\documentclass[a4paper,10pt,twoside]{cpc-hepnp}

\usepackage{multicol} 
\usepackage{graphicx}
\usepackage{booktabs}
\usepackage{amssymb,bm,mathrsfs,amscd}
\usepackage[tbtags]{amsmath}
\usepackage{lastpage}
\usepackage{psfrag}
\begin{document}
\newcommand\nn{\nonumber}
\newcommand\ba{\begin{eqnarray}}  
\newcommand\ea{\end{eqnarray}}
\newcommand{\br}[1]{\left( #1 \right)}
\newcommand{\brs}[1]{\left[ #1 \right]}
\newcommand{\brf}[1]{\left\{ #1 \right\}}
\newcommand{\brm}[1]{\left| #1 \right|}

\fancyhead[co]{\footnotesize Q.Xu~ et al: News on {\tt PHOTOS} MC: 
$\gamma^* \to \pi^+\pi^-(\gamma)$ and $K^\pm\to\pi^+\pi^-e^\pm\nu (\gamma)$}

\footnotetext[0]{Received 30 Dec. 2009}

\title{News on {\tt PHOTOS} Monte Carlo: \\
$\gamma^* \to \pi^+\pi^-(\gamma)$ and $K^\pm\to\pi^+ \pi^- e^\pm \nu(\gamma)$  
\thanks{Supported by EU Marie Curie Research Training Network grant under
the contract No. MRTN-CT-2006-0355505 and   
Polish Government grant N202 06434 (2008-2011)}}

\author{%
      Qingjun Xu $^{1, 1)}$\email{qingjun.xu@ifj.edu.pl}%
\quad Z. W\c{a}s$^{1, 2, 2)}$\email{wasm@mail.cern.ch}%
}

\maketitle

\address{%
1~(Institute of Nuclear Physics, PAN,
        Krak\'ow, ul. Radzikowskiego 152, Poland)\\
2~(CERN PH-TH, CH-1211 Geneva 23, Switzerland )}

\begin{abstract}
{\tt PHOTOS} Monte Carlo is widely used for simulating QED effects in decay 
of intermediate particles and resonances. It can be easily connected
to other main process  generators. In this paper we consider
 decaying processes 
$\gamma^* \to \pi^+\pi^-(\gamma)$ and $K^\pm\to\pi^+\pi^-e^\pm\nu$
 $(\gamma)$ in the framework of Scalar QED.
These two processes are interesting not only for the technical aspect of 
{\tt PHOTOS} Monte Carlo, but also for precision measurement
of $\alpha_{QED}(M_Z)$, $g-2$, as well as $\pi\pi$ scattering lengths.
\end{abstract}


\begin{flushright} 
{ IFJPAN-IV-2010-1} 
\end{flushright}
\begin{multicols}{2}
\section{Introduction}
In high energy experiments, one of the crucial works is to compare
new experiment results with predictions from the theory. If the
agreement is obtained, the theory is proved to be true.
Otherwise one may think that the theory calculations turned out to be wrong or 
the effect of new physics appeared. Monte Carlo generators, rather than
analytical calculations, are required to provide theoretical results 
of real experiment interest. The {\tt PHOTOS} Monte Carlo
\cite{Barberio:1990ms,Barberio:1993qi} is a universal Monte Carlo algorithm
that is designed for simulating QED radiative corrections in cascade
decays. It is widely used in low energy experiments and high energy
experiments as well. The program 
is based on exact multiphoton phase space while the matrix element 
is approximately taken as process
independent multidimensional kernel.
In some cases, the exact first order matrix element is employed in order
to improve the precision.

From the experience gained in 
KKMC project \cite{Jadach:2000ir,Jadach:1999vf}
where spin amplitudes were used,
one concludes that spin amplitudes are essential for 
design and tests of the program, in particular for choice of
the single emission kernels. 
The analysis of the spin amplitudes and tests for the algorithm in case of 
$Z$ decay into pair of
charged fermion was given in Ref. \cite{Golonka:2006tw}. 
The case of the scalar particle decay into pair of fermions is covered 
in Ref. \cite{Andonov:2002mx} and the decay of
spinless particle into pair of scalars is studied in Ref. \cite{Nanava:2006vv}. 
The case of $W$ decay was covered in Ref. \cite{Nanava:2003cg, Nanava:2009vg}.

In this paper we will study spin amplitudes for 
$\gamma^* \to \pi^+\pi^-(\gamma)$ 
decay. It not only provides
example for  studies of Lorentz and gauge group properties 
of spin amplitudes and cross sections,  
but also is important to improve 
theoretical uncertainty of {\tt PHOTOS} for this decay. Furthermore,
it is of great experiment interest since its relevance to precision 
measurement of  $\alpha_{QED}(M_Z)$ and $g-2$. 
 
$K_{e4}$ decay could give the unique information on the value of $s-$
and $p-$ wave $\pi\pi$ scattering lengths. The high statistics
measurements of $K_{e4}$ decay has been performed by NA48/2
collaboration at CERN \cite{Batley:2007zz}.  
Theoretical predictions are well calculated 
\cite{Colangelo:2001df} and the difference between theoretical prediction and
experiment value is about $7\%$.  In
particular, QED corrections to this process are known to be non-negligible. 
They need to be taken into account with the help of Monte Carlo because 
their size depend on detector acceptance. 
In NA48 experiment, to take into account QED effects,
{\tt PHOTOS} Monte Carlo is used together with Coulomb correction 
(see Ref. \cite{Batley:2007zz}).
In this paper we consider QED radiative corrections to $K_{e4}$ decay
and implement it into {\tt PHOTOS} Monte Carlo.

The paper is organized as follows:  
process $\gamma^* \to \pi^+\pi^-(\gamma)$ is studied in Section 2,
Section 3 will come to $K_{e4}$ decay, summary will be in Section 4.

\section{$\gamma^* \to \pi^+\pi^-(\gamma)$ }
In order to match the parton shower and 
the hard bremsstrahlung matrix element into interference
weight, the matrix element of $\gamma^* \to \pi^+\pi^-(\gamma)$ 
needs to be studied in great detail. Its gauge invariant parts 
need to be identified and relations to amplitudes
of lower orders have to be found. 
\subsection{Spin amplitude}
Consider the process
$e^+e^- \to \gamma^* (p) \to$\\
$ \pi^+(q_1) \pi^-(q_2) \gamma (k, \epsilon)$, see Fig. \ref{Feynman1}.
\begin{center}
\begin{tabular*}{100mm}{ll}
\includegraphics[width=3.5cm]{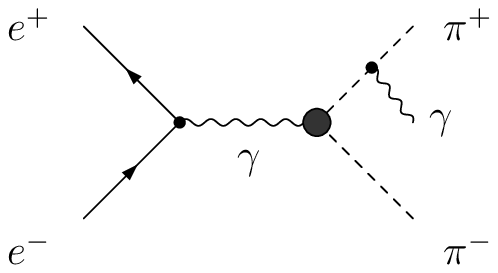}&
\includegraphics[width=3.5cm]{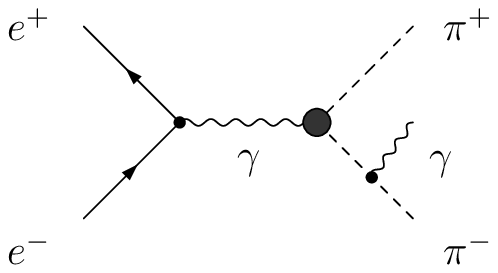}\\
\includegraphics[width=3.5cm]{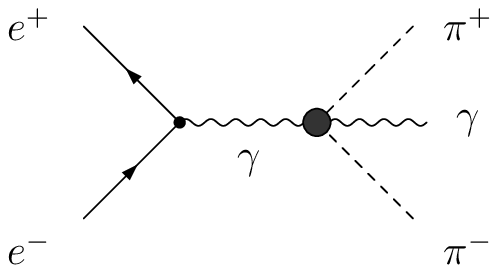} & 
\end{tabular*}
\figcaption{Feynman diagrams of $e^+e^- \to \gamma^*\rightarrow 
\pi^+\pi^- \gamma$\label{Feynman1}}
\end{center}
The amplitudes equivalent to those given in Ref. \cite{Czyz:2002np} 
are obtained.
It can be written as 
\begin{equation}
M=V^\mu H_\mu, 
\end{equation}
where  
$V_{\mu} =\bar{v}(p_1,\lambda_1) \gamma_\mu u(p_2,\lambda_2)$. The
$p_1,\lambda_1,p_2,\lambda_2$ are momenta and helicities of the incoming
electron and positron. 
 $V_{\mu}$ defines the spin state of the intermediate $\gamma^*$.

Let us focus on the part for virtual photon decay. Following conventions
of Ref. \cite{Czyz:2002np}, the final interaction part of the
Born matrix element for such process is 
\begin{equation}
H_0^\mu(p,q_1,q_2) = \frac{eF_{2\pi}(p^2)}{p^2}(q_1-q_2)^{\mu}.
\end{equation}
Here $p=q_1+q_2$. If photon is present, this part 
of the amplitude reads:
\begin{eqnarray}\label{eq:hmu1}
H^\mu &=&\frac{e^2F_{2\pi}(p^2)}{p^2}\left \{(q_1+k-q_2)^{\mu}
\frac{q_1\cdot \epsilon^*}{q_1\cdot k} +\right. \nonumber \\
&& \left. (q_2+k-q_1)^{\mu}
\frac{q_2\cdot \epsilon^*}{q_2\cdot k}-2\epsilon^{*\mu} \right\}.
\end{eqnarray}
It makes sense to rewrite
Eq. (\ref{eq:hmu1}) explicitly as sum of two gauge invariant terms:
\begin{equation}
H^\mu =H^\mu_{I} +H^\mu_{II}, 
\label{AmpliS}
\end{equation}
\begin{eqnarray}
H_{I}^\mu &=&\frac{e^2F_{2\pi}(p^2)}{p^2}\left ((q_1-q_2)^{\mu}+k^{\mu}
\frac{q_2\cdot k - q_1\cdot k}{q_2\cdot k + q_1\cdot k} \right)
\nonumber \\
&& \left (\frac{q_1\cdot \epsilon^*}{q_1\cdot k} - 
\frac{q_2\cdot \epsilon^*}{q_2\cdot k}\right ), 
\label{HIp}
\end{eqnarray}
\begin{eqnarray}
H_{II}^{\mu}=\frac{2e^2F_{2\pi}(p^2)}{p^2}
\left (\frac{k^{\mu}(q_1\cdot \epsilon^*+ q_2\cdot \epsilon^*)}
{q_2\cdot k + q_1\cdot k} - \epsilon^{*\mu}\right )\, .
\label{HIIp}
\end{eqnarray}
One can easily see that Eq.(\ref{HIp}) has a typical form for amplitudes
of QED exclusive exponentiation \cite{Jadach:1998jb}, that is Born-like 
-expression multiplied by an eikonal factor 
$\left (\frac{q_1\cdot \epsilon^*}{q_1\cdot k} - 
\frac{q_2\cdot \epsilon^*}{q_2\cdot k}\right )$. The expression in front 
of the factor indeed 
approaches the Born one in soft photon and collinear photon limit.
Thus, it is consistent with LL level factorization into
Born amplitude and eikonal factor.

If one takes separation (\ref{AmpliS}) for the calculation of two parts
of spin amplitudes, then after spin average, the expression for the 
cross section takes the form:
\begin{eqnarray}
\sum_{\lambda,\epsilon} |M|^2 = \sum_{\lambda,\epsilon} |M_{I}|^2
+ \sum_{\lambda,\epsilon} |M_{II}|^2 + 
2\sum_{\lambda,\epsilon} M_{I}M_{II}^*\, .
\label{Mone}
\end{eqnarray}
We should stress that Eq.(\ref{Mone}) 
can have its first term even closer to Born-times-eikonal-factor 
form. For that purpose it is enough to adjust normalization of the first part
of Eq.(\ref{Mone}) to Born amplitude times eikonal factor, and replace 
$|M_{I}|^2$ with 
\begin{eqnarray}
|M_{I}^\prime|^2 = |M_I|^2 \frac{|\vec {q_1} - \vec {q_2}|^2_{\rm Born}}
{|\vec {q_1} - \vec {q_2} + \vec k \frac{q_2\cdot k-q_1 \cdot k}
{q_2\cdot k+q_1 \cdot k}|^2}\, . 
\end{eqnarray}
Compensating adjustment to the remaining parts of Eq.(\ref{Mone}) is then necessary. 
%The change can be performed by numerical manipulation of the three parts of 
%Eq.(\ref{Mone}). 
Since $\sum_{\lambda,\epsilon} |M_{I}^\prime|^2$ 
is the expression used in {\tt PHOTOS} Monte Carlo in 
Ref. \cite{Nanava:2006vv}, 
such a modification is of interest.
In the next section,
we will perform our numerical investigations with 
respect to Ref. \cite{Nanava:2006vv} which is a reference for us. 

\subsection{Numerical results}
We will show results at 2 GeV center of mass energy. 
Comparison of result from $ \sum_{\lambda,\epsilon} |M_{I}^\prime|^2$
with result from {\tt PHOTOS} 
with matrix element taken from Ref. \cite{Nanava:2006vv} is shown
in Fig.\ref{fig:one}. 
One can see that agreement is excellent
all over the phase space. 
It is true only for the case when distributions are averaged 
over the orientation of the whole event with respect to incoming beams 
(or spin state of the virtual photon). 

If instead of $\sum_{\lambda,\epsilon} |M_{I}^\prime|^2$
one would use directly $\sum_{\lambda,\epsilon} |M_{I}|^2$,
that is when normalization of 
Born-like factor is not performed, difference with respect to formulas
in Ref. \cite{Nanava:2006vv} is much larger, see Fig. \ref{fig:two}.
\begin{center}
\psfrag{M2pipi}{\tiny \bf $M_{\pi+\pi-}^2/S$}
\includegraphics[width=8cm]{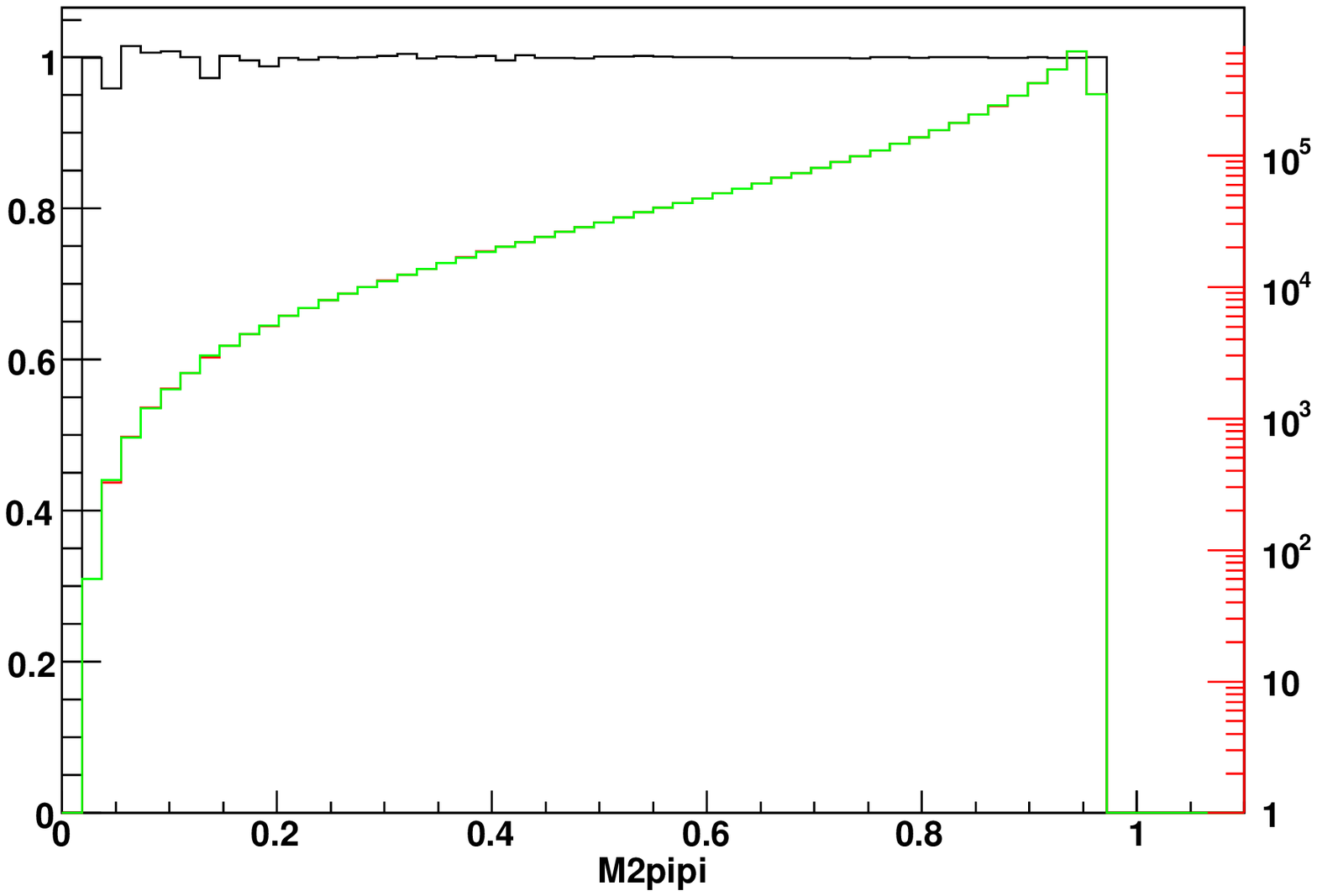}
\figcaption{\label{fig:one}Comparison of result using 
$ \sum_{\lambda,\epsilon} |M_{I}^\prime|^2$(green line) with that 
using  matrix element taken from Ref. \cite{Nanava:2006vv} (red line).
Black line represents their ratio.}
\end{center}
%Note tht fraction of events with photons above
%50 MeV is respectively $4.2279 \pm 0.0021 \%$ and $4.2269 \pm 0.0021\%$ 
%for these two programs.
\begin{center}
\psfrag{M2pipi}{\tiny \bf $M_{\pi+\pi-}^2/S$}
\includegraphics[width=8cm]{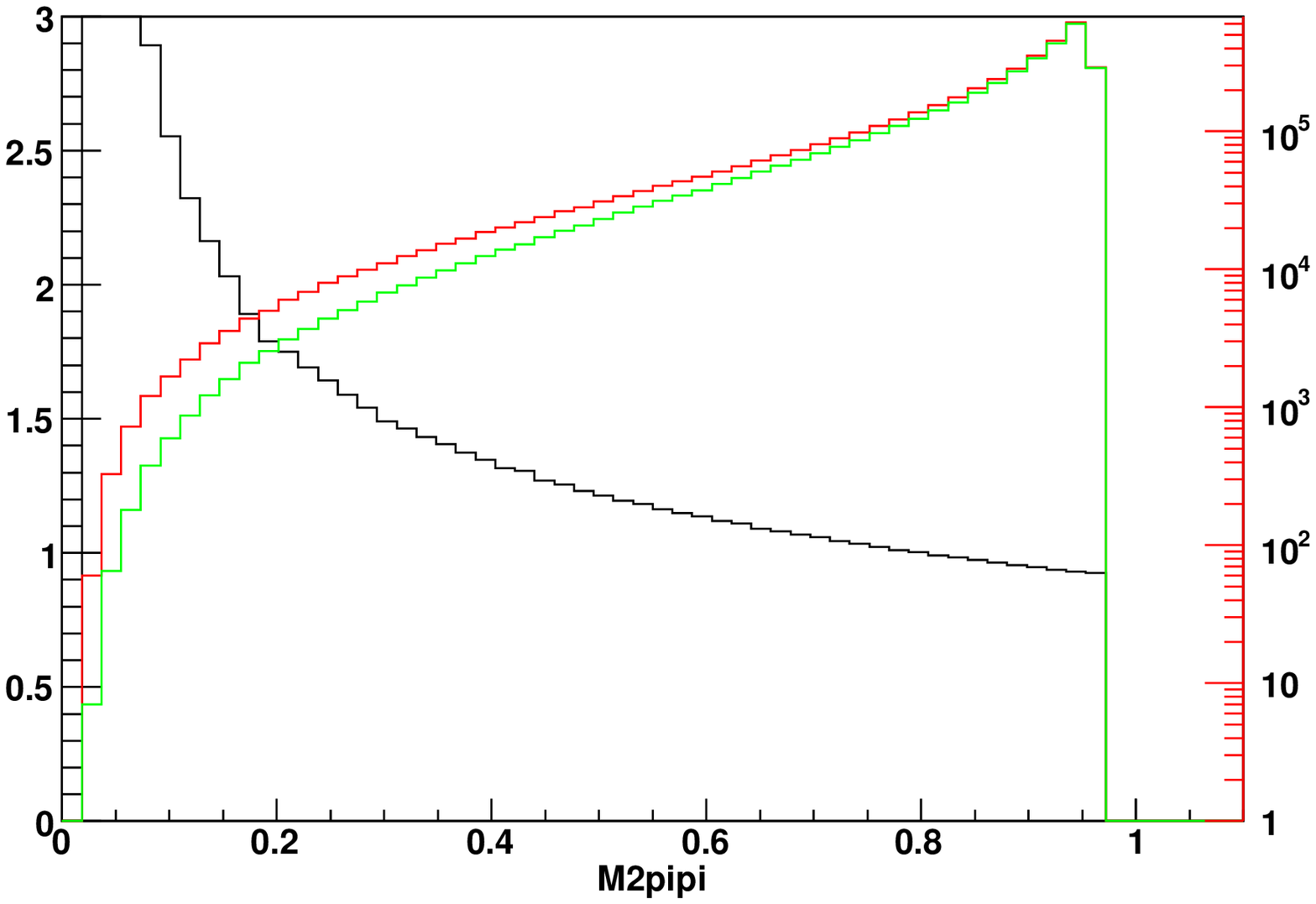}
\figcaption{\label{fig:two}Comparison of result using
$ \sum_{\lambda,\epsilon} |M_{I}|^2$(green line) with that 
using matrix element taken from Ref. \cite{Nanava:2006vv} (red line).
Black line represents their ratio. }
\end{center}

Finally let us compare result of complete scalar QED matrix element 
with that of matrix element taken from Ref. \cite{Nanava:2006vv}, 
see Fig.~\ref{fig:four}. 
At high photon energy   
\begin{center}
\psfrag{M2pipi}{\tiny \bf $M_{\pi+\pi-}^2/S$}
\includegraphics[width=8cm]{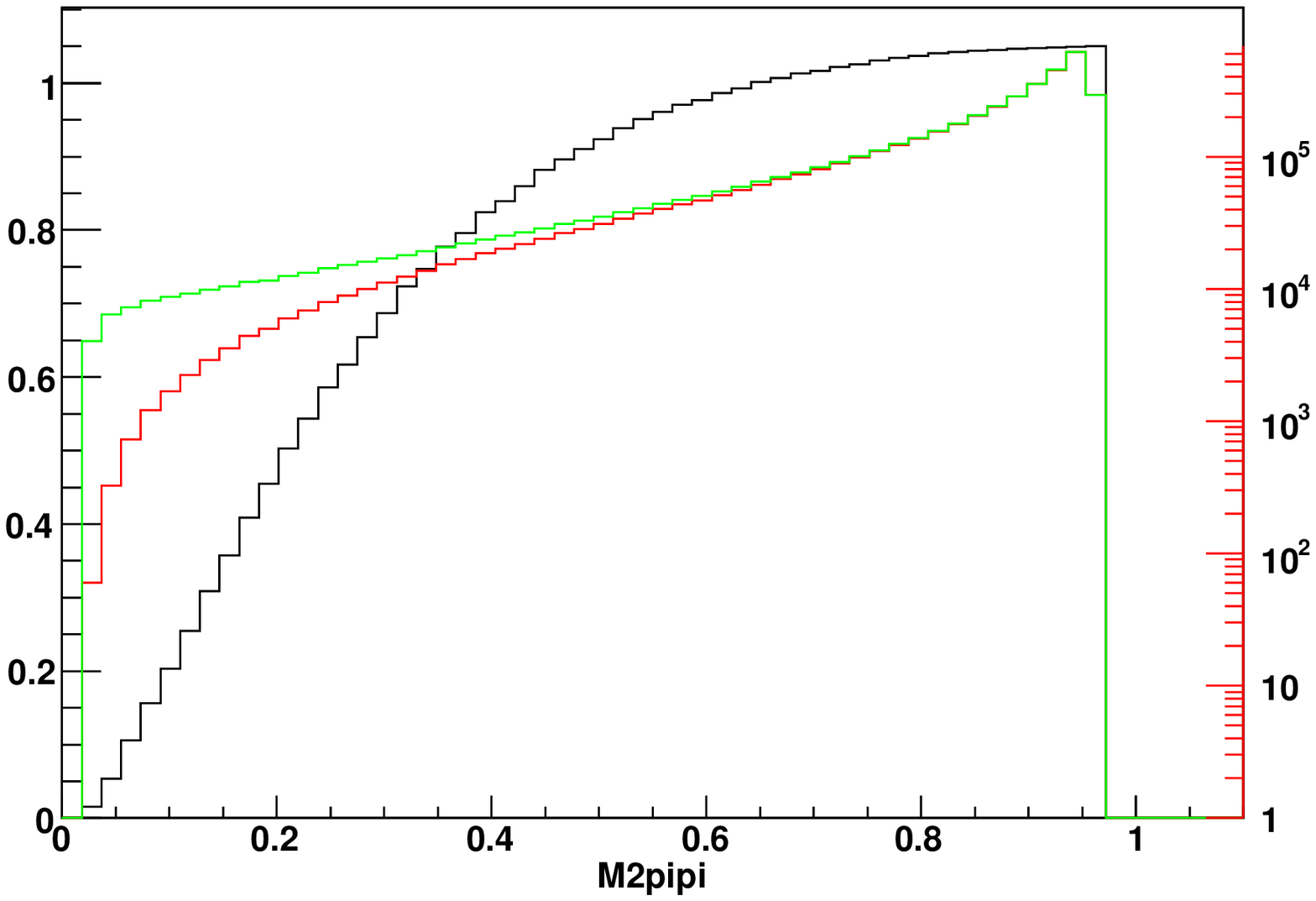}
\figcaption{\label{fig:four}Comparison of result using
complete matrix element (green line) with that 
using matrix element taken from Ref. \cite{Nanava:2006vv} (red line).
Black line represents their ratio. }
\end{center}
region, there is clear surplus of events 
with respect to formula in
Ref. \cite{Nanava:2006vv}. That contribution should not be understood 
as bremsstrahlung, but rather as genuine process. 
Anyway in that region of phase space scalar QED is not expected to work well. 
Note that the difference between results of Figs. 
 \ref{fig:one} and  \ref{fig:four} is only $0.2 \%$ of the total process 
rate. That is why our detailed discussion is not important for numerical
conclusions, but important for understanding the underlying structure.

From Fig. \ref{fig:one} one could conclude that the universal kernel 
in Ref. \cite{Nanava:2006vv}, for arbitrary large samples,
is equivalent to the matrix element 
 $ \sum_{\lambda,\epsilon} |M_{I}^\prime|^2$. But differences appear in
distributions sensitive to initial state spin 
orientation, see Figs. \ref{fig:oneX} and \ref{fig:oneY}. 
On these plots angular distributions of the photon momentum and 
$\pi^+$ momentum with respect
to the beam line are shown, respectively.
Regions of phase space giving near zero contribution at the Born level
are becoming more populated if approximation for the photon radiation 
matrix element \cite{Nanava:2006vv} is used. 
\begin{center}
\psfrag{cosThetaGamma}{\scriptsize $\cos\theta_{\gamma}$}
\includegraphics[width=8cm]{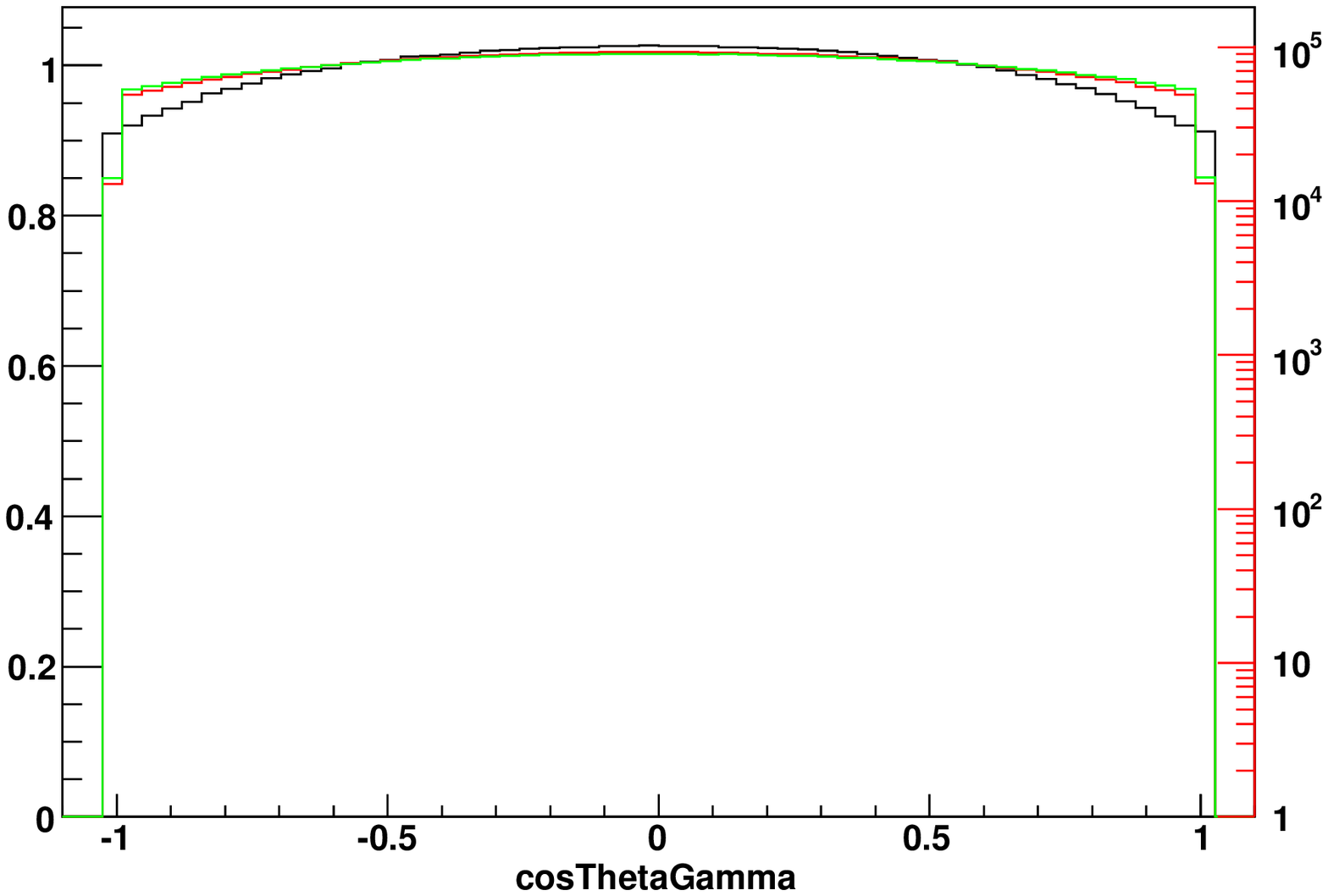}
\figcaption{\label{fig:oneX}Comparison of result using
$ \sum_{\lambda,\epsilon} |M_{I}^\prime|^2$(green line) with that 
using matrix element taken from Ref. \cite{Nanava:2006vv} (red line).
Black line represents their ratio. }
\end{center}
\begin{center}
\psfrag{cosThetaPiPlus}{\scriptsize $\cos\theta_{\pi+}$}
\includegraphics[width=8cm]{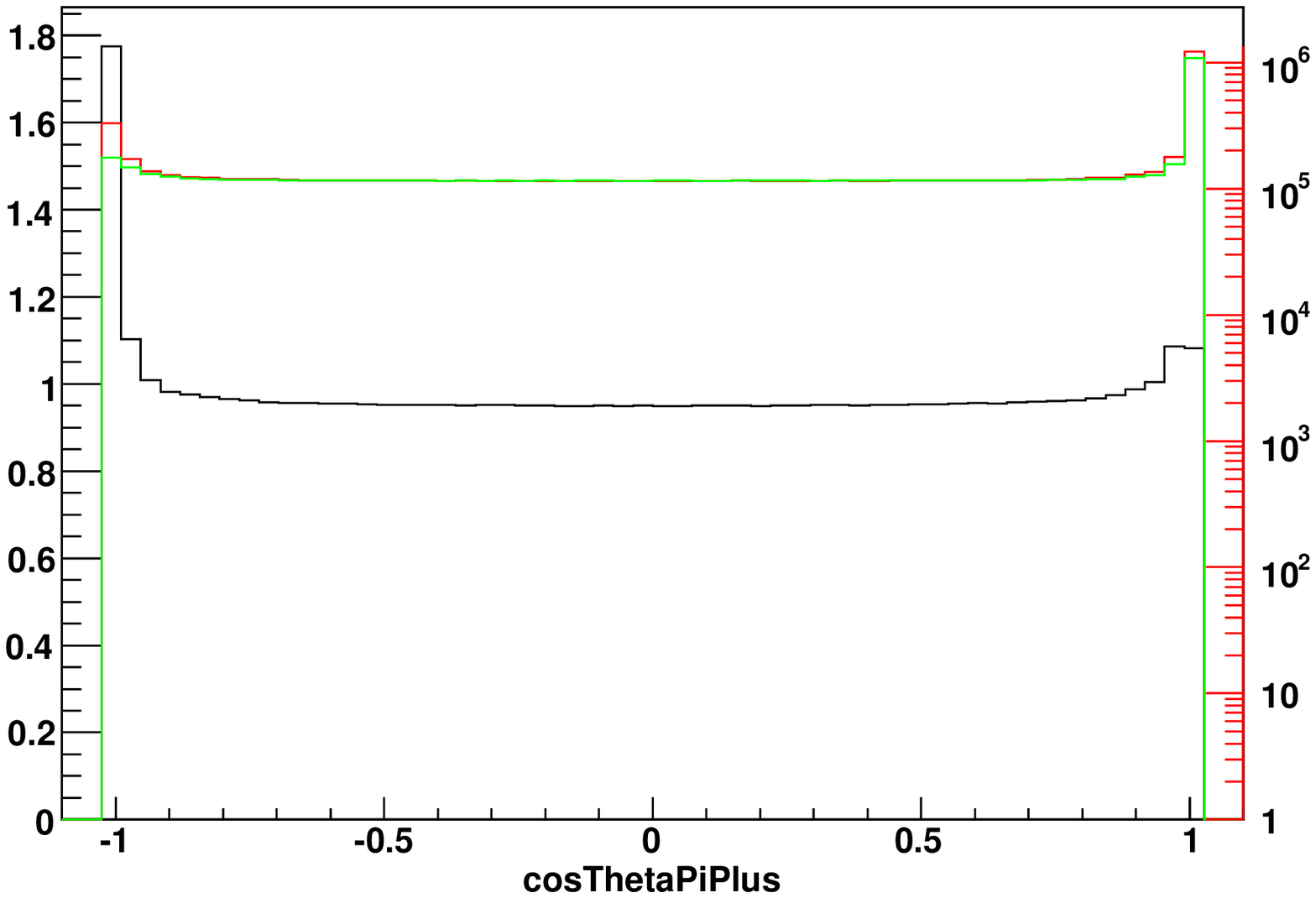}
\figcaption{\label{fig:oneY}Comparison of result using
$ \sum_{\lambda,\epsilon} |M_{I}^\prime|^2$(green line) with that 
using matrix element taken from Ref. \cite{Nanava:2006vv} (red line).
Black line represents their ratio. }
\end{center}
\section{$K^\pm\to\pi^+ \pi^- e^\pm \nu (\gamma)$}
In this section radiative corrections to $K_{e4}$ decay are calculated
following approximations explained in Ref. \cite{Bystritskiy:2009iv1}.  
We compare it with Coulomb corrections used by NA48/2 collaboration.
The hard photon bremsstruhlung is calculated analytically with 
soft photon and collinear photon approximation. Numerical tests with  
{\tt PHOTOS} Monte Carlo are performed. 
\subsection{QED radiative corrections to $K_{e4}$ decay}
Consider $K_{e4}$ decay, 
\ba
 K^\pm(p)\to\pi^+(q_+)+\pi^-(q_-)+e^\pm(p_e)+\nu(p_\nu)\, .
 \label{Born}
\ea
%\begin{center}
%\includegraphics[width=8cm]{FigBorn.eps}
%\figcaption{Born diagram of $K_{e4}$ decay}
%\end{center}
In the framework of Scalar QED but neglecting diagrams  
with photons emission from hadronic or weak blocks,
one can calculate the virtual photon corrections. 
Note that the electron photon vertex is taken from standard QED. 
Contribution of virtual diagrams reads 
\begin{center}
\includegraphics[width =8cm]{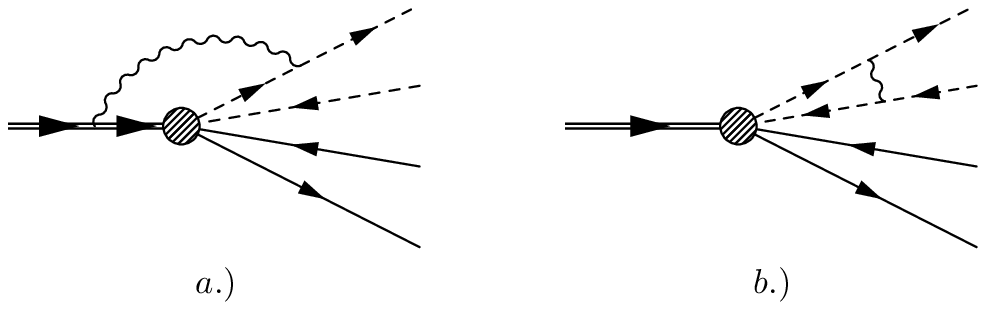}
\figcaption{example vertex diagrams \cite{Bystritskiy:2009iv1}}
\end{center}
\ba
\frac{d\Gamma_{\rm virt}}{d\Gamma_{\rm Born}}&=&\frac{\alpha}{\pi}
  \left[\ln \frac{m}{\lambda}\left (4+\frac{L_-}{\beta_-}-\frac{L_+}{\beta_+}
    -2\rho -\frac{1+\beta^2}{\beta}L_\beta\right. \right. \nn \\
&& \left. \left. 
+2\ln{\frac{p_e\cdot q_+}{p_e\cdot q_-}}\right )+ 
\pi^2\frac{1+\beta^2}{2\beta}+\rho^2+\frac{1}{2}\rho\right. \nn \\
& & \left. + 2\rho \ln
\frac{m}{2E_e}+\frac{9}{4}\log\frac{\Lambda^2}{m^2}+K_v
\right],
\label{VirtRC}
\ea
where $m$ is the charged pion mass, $\lambda$ is 
photon mass used as infrared regulator, 
$\Lambda$ is ultraviolet (UV) cut off.  
In Eq.(\ref{VirtRC}) we have defined
\ba
\rho&=&\ln\frac{2E_e}{m_e}, \nn \\
\beta &=&\sqrt{1-\frac{4m^2}{s_\pi}},\qquad
 L_\beta =\ln\frac{1+\beta}{1-\beta}, \nn\\
\beta_\pm &=& \sqrt{1-\frac{m^2}{E_\pm^2}},\qquad
L_\pm =\ln\frac{1+\beta_\pm}{1-\beta_\pm}. 
\ea
$K_v$ depends on masses of particles and kinematics. 
Note that the UV-divergent part in Eq.(\ref{VirtRC}) will cancel 
if the renormalized coupling $G_F^2 V_{us}^2$ is used 
instead of bare ones, 
\ba
G_F^2 V_{us}^2 \to\left (1 -\frac{9}{4}\log\frac{\Lambda^2}{m^2}\right ) 
\left (G_F^2 V_{us}^2\right)^{\rm bare}\, .
\label{Recoupling}
\ea

The soft photon contribution can be easily obtained, resulting in 
expression which is gauge invariant: 
\ba
\frac{d\Gamma_{\rm soft}}{d\Gamma_{\rm Born}}=
-\frac{\alpha}{4\pi^2}\int\frac{d^3k}{\omega}
\br{\frac{p}{pk}+\frac{q_-}{q_-k}-\frac{q_+}{q_+k}-\frac{p_e}{p_ek}}^2,
\label{SoftIntegration}
\ea
here $k$ is photon momentum, $\omega$ is photon energy. 
It is straightforward to integrate out solid angle of photon 
momentum $k$ and over its energy $\omega$ up to a limit 
 $\omega < \Delta \epsilon$; we obtain:
\ba
\frac{d\Gamma_{\rm soft}}{d\Gamma_{\rm Born}}&=&
\frac{\alpha}{\pi}\left [\ln\br{\frac{2\Delta\epsilon}{\lambda}}
\left (-4-\frac{L_-}{\beta_-}+\frac{L_+}{\beta_+}\right. \right.
\nn \\
& & \left.\left. 
 +2\rho+\frac{1+\beta^2}{\beta}L_\beta-2
\ln{\frac{2p_e\cdot q_+}{2p_e\cdot q_-}}\right )
\right.\nn \\
&& \left. 
+\rho-\rho^2+ K_s \right ].
\label{SoftRC}
\ea
Function $K_s$ is dependent on masses of particles and kinematics.
Soft singularity is again regularized with the photon mass $\lambda$.

The contribution of soft and virtual photons can be easily combined.
It reads
\ba
\frac{d\Gamma_{\rm Born + virt + soft}}{d\Gamma_{\rm Born}}
&=&1 + \sigma P_\delta +
\frac{\pi \alpha (1+ \beta^2)}{2\beta} \nn \\
&&  +\frac{\alpha}{\pi}K_{vs}\, ,
\label{BVSRC}
\ea
where 
\ba
P_\delta&=&2\ln\frac{\Delta \epsilon}{E_e}+\frac{3}{2},  \quad
\sigma=\frac{\alpha}{2\pi}(2 \rho -1), 
\ea
the expression of $K_{vs}$ depends not only on  
masses of particles and kinematics, but also on soft photon energy cutoff
$\Delta\epsilon $.   

Staring from a certain energy threshold photons can be observed, 
at least in principle. Such 
contribution is called hard photon radiation and we will assume 
that it is defined by condition
$\omega > \Delta \epsilon$. 
If soft photon matrix element Eq.(\ref{SoftRC}) is used, and 
 maximum energy of photon is taken from kinematical constraint 
$E_e$, then Eq. (\ref{SoftRC}) can be used to 
calculate leading double logarithmic result of hard photon radiation.
For that reason
one need to subtract Eq.(\ref{SoftRC})
calculated for $E_e$ from that for $\Delta \epsilon$. 
Thus the contribution of hard photon emission by 
soft-photon-approximation is obtained, 
\ba
\frac{d\Gamma_{\rm Hard_{soft-like}}}{d\Gamma_{\rm Born}}&=&
\frac{\alpha}{\pi}\ln\br{\frac{\Delta\epsilon}{E_e}}
\left (4+\frac{L_-}{\beta_-}-\frac{L_+}{\beta_+}-2 \rho \right. \nn \\
& & \left. -\frac{1+\beta^2}{\beta}L_\beta+
2\ln\frac{2p_e\cdot q_+}{2p_e\cdot q_-}
\right).
\label{soft-approx}
\ea
This result has to be corrected for single logarithm related to
collinear photon emission along the charged outgoing decay products.

Photon emission from $e^\pm$ will give collinear singularity.
Remaining charged products, $\pi^+$ and $\pi^-$ are not relativistic.
The electron  part can be calculated with the help of
collinear-photon-approximation. 
The squared collinear amplitude can be found in Ref. \cite{baier:1973ms}. 
The actual formula for collinear contribution is 
taken from Ref. \cite{xu:2006thesis}, 
\ba
\frac{d\Gamma_{\rm Collinear}}{d\Gamma_{\rm Born}}= 
\frac{\alpha}{4\pi^2} \int \frac{d^3k}{\omega}\frac{z^2}{p_e \cdot k} 
\left (\frac{1+z^2}{1-z} - \frac{m_e^2}{p_e \cdot k}\right),
\ea
here $z = \frac{E_e}{E_e + \omega}$. 
First we integrate over solid angle of photon direction, 
later we integrate 
photon energy from $\Delta \epsilon$ to the maximum value
$E_e$. Then the result of hard photon emitting from $e^\pm$ is obtained 
using collinear-photon-approximation, 
\ba
\frac{d\Gamma_{\rm Hard_{e}}}{d\Gamma_{\rm Born}}&=&
- \sigma P_\delta + \frac{\alpha}{2\pi}\left (3 - \frac{2}{3}\pi^2
\right).
\label{colli-approx}
\ea
Eq. (\ref{colli-approx}) includes term 
proportional to $2\rho -1$ which is already present 
in Eq.(\ref{soft-approx}). This would lead to double counting. That is why
we remove this term from Eq.(\ref{soft-approx}) and obtain
 \ba
\frac{d\Gamma_{\rm Hard_{no-e}}}{d\Gamma_{\rm Born}}&=&
\frac{\alpha}{\pi}\ln\br{\frac{\Delta\epsilon}{E_e}}
\left (3+\frac{L_-}{\beta_-}-\frac{L_+}{\beta_+} \right. \nn \\
&& \left. 
- \frac{1+\beta^2}{\beta}L_\beta+
2\ln\frac{2p_e\cdot q_+}{2p_e\cdot q_-}\right).
\ea
Finally hard (real)  photon bremsstrahlung for photons of energy above 
${\Delta\epsilon}$ reads
\ba
\frac{d\Gamma_{\rm Hard}}{d\Gamma_{\rm Born}} & = &
\frac{d\Gamma_{\rm Hard_{e}}}{d\Gamma_{\rm Born}} + 
\frac{d\Gamma_{\rm Hard_{no-e}}}{d\Gamma_{\rm Born}}.
\label{hard}
\ea

If one adds real and virtual photons contribution together, one can obtain 
the expression as following,
\ba
\frac{d\Gamma_{\rm Born + virt + real}}{d\Gamma_{\rm Born}}= 1 + 
\frac{\pi \alpha (1+ \beta^2)}{2\beta} + \frac{\alpha}{\pi}K.
\label{TotalRC}
\ea
Complicated, but numerically small function  $K$ is 
dependent on masses of particles and kinematics of this process. 
Note that our final analytical result does not depend on 
the large logarithm $\ln\frac{2E_e}{m_e}$, 
or soft photon energy cut $\Delta\epsilon$. 

As one can see, Eqs. (\ref{BVSRC}) (\ref{hard}), (\ref{TotalRC})
are obtained with the help of approximations. Effectively it was assumed that 
matrix element at Born level can be always factorized out and photonic corrections can be calculated independently. Further 
corrections are assumed to be negligible and not affecting the substantial
nature properties 
of hard interaction. This may be good as starting point, but cannot be 
left without future discussion/improvements\footnote{We are grateful to
Prof. J. Gasser for stressing this point.}.

Our formulas are based on the same scheme of calculation as explained in
Ref. \cite{Bystritskiy:2009iv1} and in principle 
they should coincide numerically. 
Some differences in analytical results are still present. Also
some numerical results still remain different. 
The exact expressions for $K_s$, $K_v$, $K_{vs}$ and $K$,
as well as differences 
between our analytical results and these in Ref.\cite{Bystritskiy:2009iv1}
will not be listed here for the limit of paper length. They will be present
elsewhere.
\subsection{Numerical tests}
Let's start with Eq.(\ref{TotalRC}), the correction 
for distribution when photon is integrated out but all
other kinematic variables are kept. 
In Figs. \ref {fig:coulomb} and \ref{fig:totalRC} we show
that dominant part of 
Eq.(\ref{TotalRC}) represents Coulomb correction. The difference 
 is much smaller 
\begin{center}
\includegraphics[width=7cm]{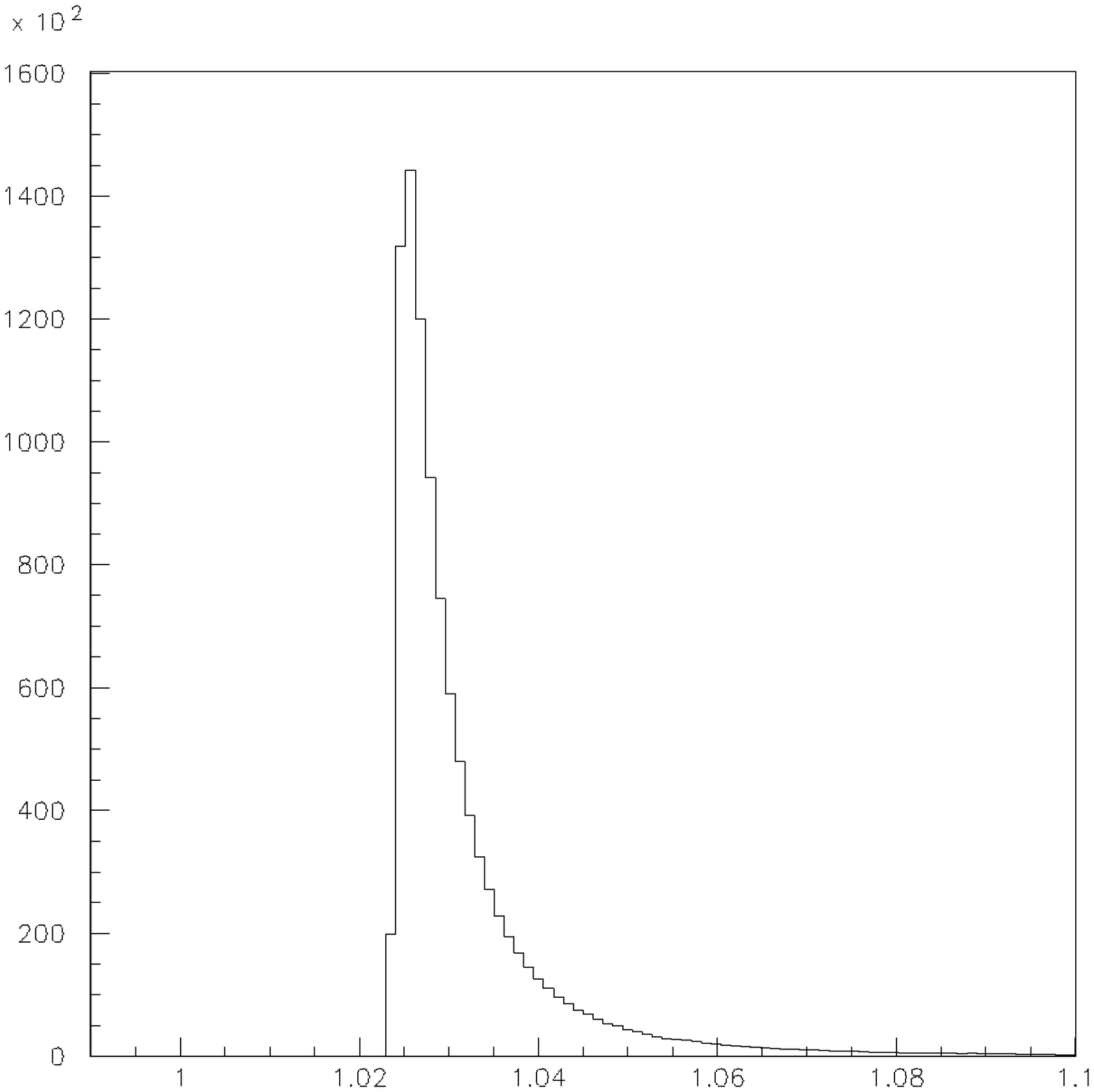}
\figcaption{\label{fig:coulomb}Coulomb correction from
 Ref. \cite{Batley:2007zz}.}
\end{center}
\begin{center}
\includegraphics[width=7cm]{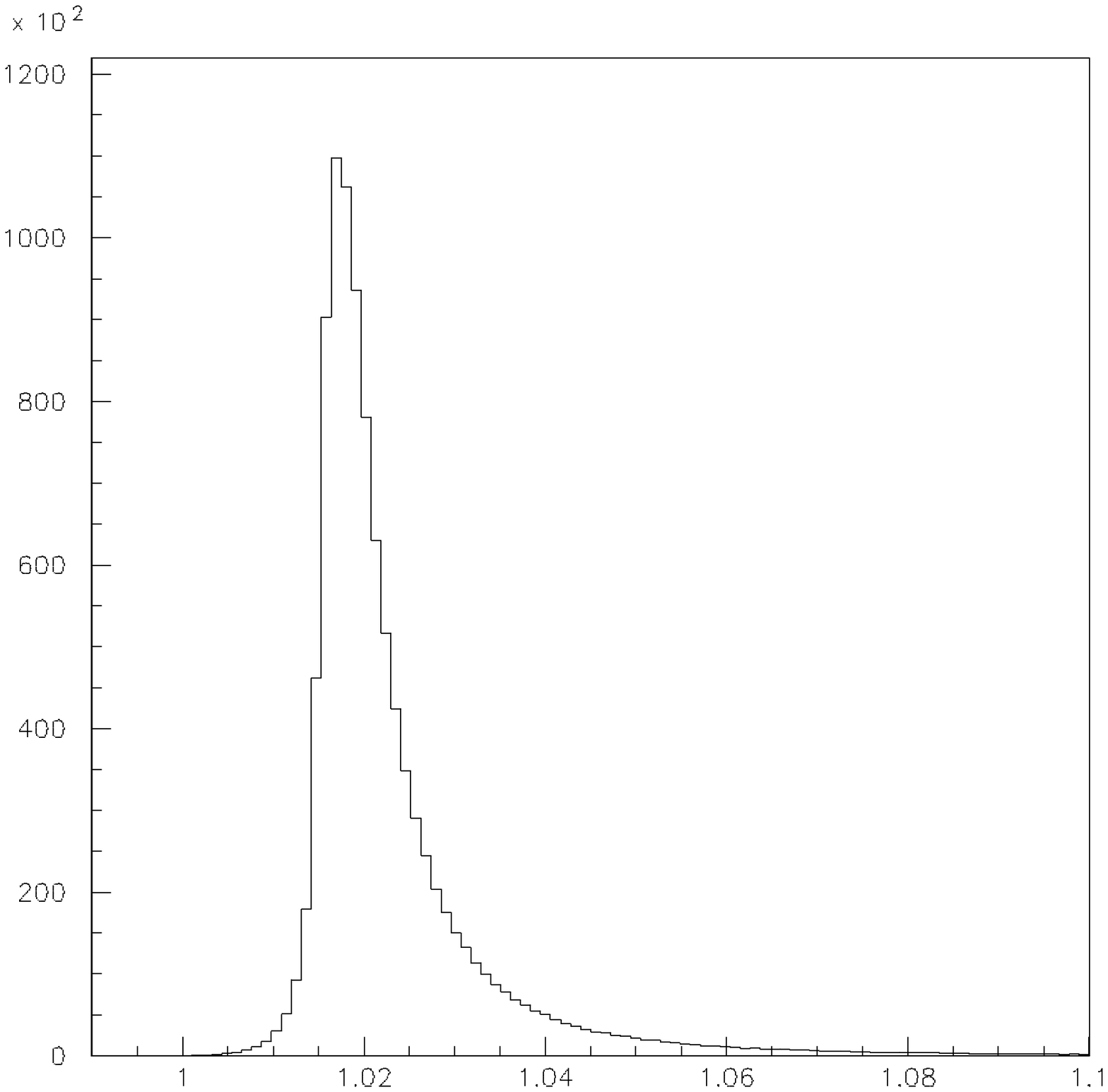}
\figcaption{\label{fig:totalRC}Radiative correction in Eq. (\ref{TotalRC}).}
\end{center}
than the effect of Coulomb correction
itself, see Fig.\ref {fig:diffcoulomb} where results for 1000 000 
Born level events are placed in the histograms.
We may conclude that our numerical implementation of Eq.(\ref{TotalRC})
works well since its dominant part represents Coulomb correction.
 \begin{center}
\includegraphics[width=7cm]{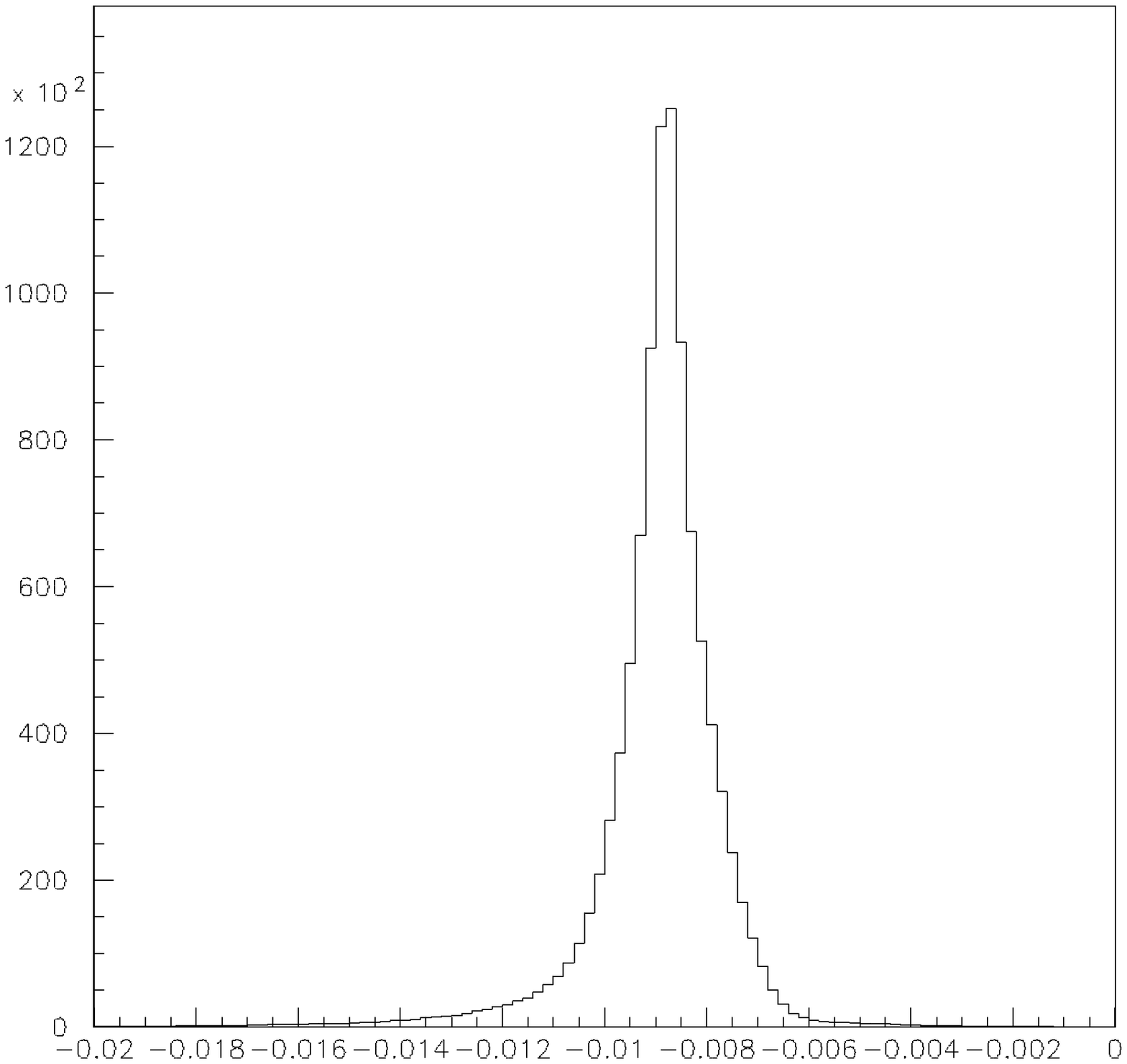}
\figcaption{\label{fig:diffcoulomb}Difference between radiative correction 
in Eq. (\ref{TotalRC}) and Coulomb correction from Ref.\cite{Batley:2007zz}
calculated event by event.}
\end{center}  

We have done numerical tests with {\tt PHOTOS} for soft photons and found
the distribution of soft photons from {\tt PHOTOS} is as same as given 
in the soft photon expression (Eq.(\ref{SoftIntegration})). 
We also have done the similar test using hard photon expression 
(Eq. (\ref{hard})) and found it matches result from  {\tt PHOTOS} simulation
in the soft photon region, too, as it is expected.  

For harder photons, we compare result of {\tt PHOTOS}, with results of 
Eq.(\ref{SoftRC}) for soft photon and Eq.(\ref{hard}) for hard photon.
Their ratios are given  
in Fig.~\ref{check_soft-hard_PHO}, where effect as function of
upper limit on photon energy was used. 100 000 000  {\tt PHOTOS} event
samples were generated using single fixed Born level event. 
\begin{center}
\includegraphics[width=8cm]{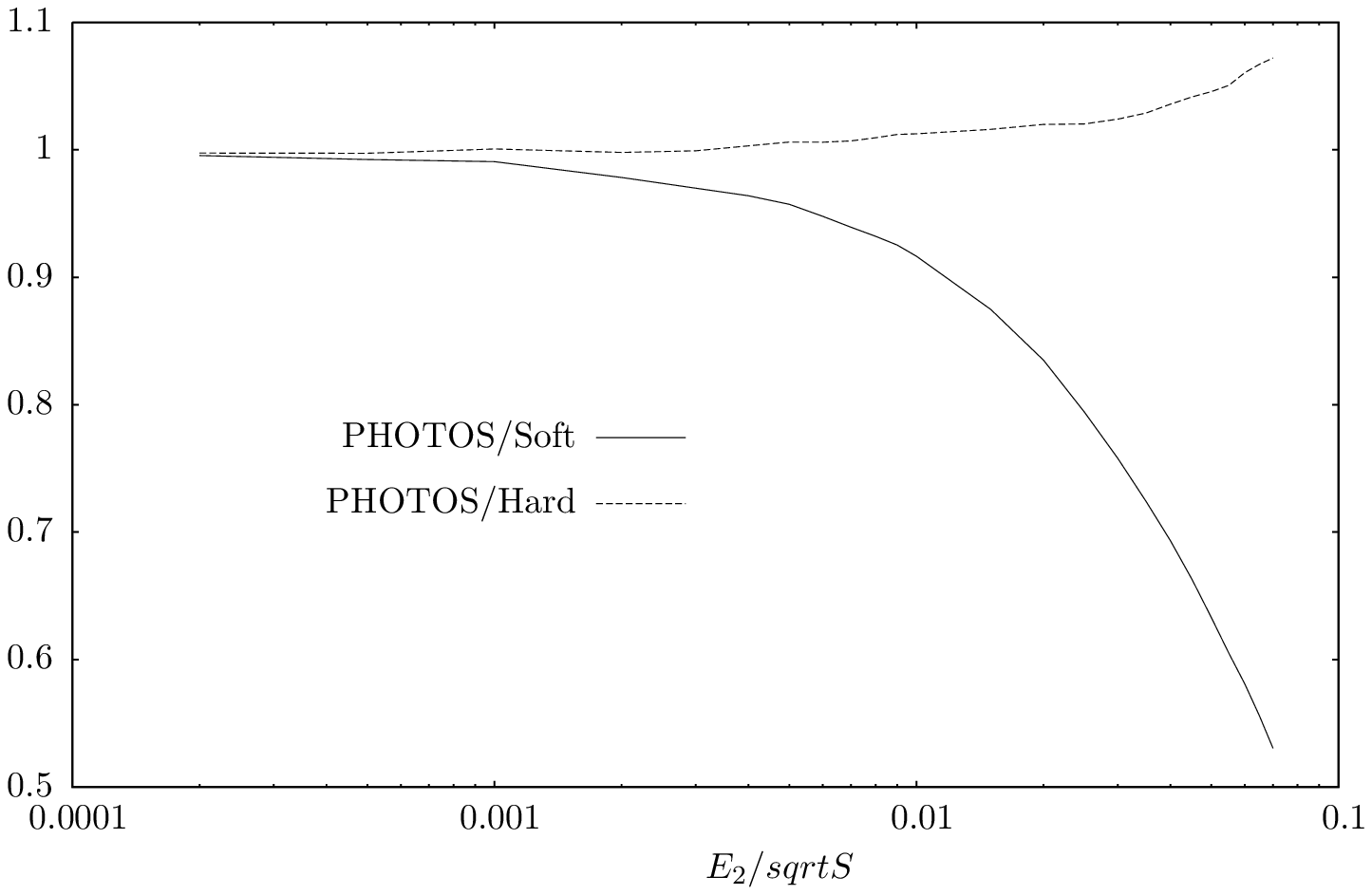}
\figcaption{\label{check_soft-hard_PHO}Ratios of result from  {\tt PHOTOS} to 
results using soft photon Eq. (\ref{SoftRC}) and 
hard photon Eq. (\ref{hard}). Here $E_1 < E_{\gamma}<E_2$ where we set $\frac{E_2}{E_1}=2.5$, $E_2$ varies
 from $0.0002\sqrt S$ to $0.07\sqrt S$. Note that the energy of electron
 $E_e \approx 0.07\sqrt S$.}
\end{center} 

We can conclude that agreement is good, as expected. Though differences 
especially in harder photon energy ranges can be seen. 
However even at the end of the spectrum, where distribution is poorly
populated, differences are at 10 \% level only.
\section{Summary}

We have presented the new tests of {\tt PHOTOS} Monte Carlo, where 
the exact matrix element of $\gamma^* \to \pi^+\pi^- \gamma$ is implemented 
and its numerical result is compared with the kernel of {\tt PHOTOS}.
%Studying spin amplitude of this process can be used as building 
%brick for amplitudes of more elaborated processes.  
QED radiative correction to process $K^\pm \to l^\pm \nu \pi^+ \pi^- (\gamma)$  
is also studied analytically. Reasonable numerical agreement with simulations 
including Coulomb correction and
{\tt PHOTOS} Monte Carlo was found. 
Since several assumptions are employed,
further work is necessary.
Our result is of practical interest for experiments. 
They confirm that at least on technical level 
the Monte Carlo program works well as expected.

\centerline{\bf Acknowledgments}

Useful and inspiring discussions with B. Bloch and J. Gasser are acknowledged.
We acknowledge also discussions with E. Kuraev.

\end{multicols}

\clearpage

\end{document}